\RequirePackage[l2tabu, orthodox]{nag}

\PassOptionsToPackage{unicode, pdfa}{hyperref}
\PassOptionsToPackage{table, hyperref}{xcolor}

\documentclass [
	a4paper     ,
	USenglish   ,
	pdfa        ,
	cleveref    ,
	autoref     ,
	thm-restate ,
	numberwithinsect,
	]{lipics-v2021}
\pdfoutput=1

\hideLIPIcs{}
\nolinenumbers{}

\usepackage{wfc-preamble}
\usepackage{wfc-commands}

\graphicspath{{assets/}}

\title{Fault Tolerant Coloring of the Asynchronous Cycle}

\keywords{%
	graph coloring, LOCAL model, shared-memory model, immediate snapshot, renaming, wait-free algorithms.
}

\ccsdesc{Theory of computation~Distributed algorithms}
\ccsdesc{Mathematics of computing~Graph coloring}
\ccsdesc{Computer systems organization~Dependable and fault-tolerant systems and networks}
\ccsdesc{Theory of computation~Models of computation}

\author{Pierre Fraigniaud}
 	{Université Paris Cité, CNRS, IRIF, F-75013, Paris, France}
 	{pierre.fraigniaud@irif.fr}
 	{0000-0003-4534-4803}{additional support from the project ANR-20-CE48-0006 (\textsc{ducat}).}

\author{Patrick Lambein-Monette}
	{Université Paris Cité, CNRS, IRIF, F-75013, Paris, France}
	{patrick.lambein@irif.fr}
	{0000-0002-9401-8564}{}

\author{Mikaël Rabie}
	{Université Paris Cité, CNRS, IRIF, F-75013, Paris, France}
	{mikael.rabie@irif.fr}
	{0000-0001-6782-7625}{}

\authorrunning{P. Fraigniaud, P. Lambein-Monette and M. Rabie}
\Copyright{P. Fraigniaud et al.}

\begin{document}
\maketitle

\begin{abstract}
We present a wait-free algorithm
	for proper coloring the $n$~nodes of the asynchronous cycle~$C_n$,
	where each crash-prone node starts with its (unique) identifier as input.
The algorithm is independent of~$n \geq 3$,
	and runs in $\BigO{\log^*n}$ rounds in~$C_n$.
This round-complexity is optimal thanks to a known matching lower bound,
	which applies even to synchronous (failure-free) executions.
The range of colors used by our algorithm, namely $\Set{0,\dots,4}$,
	is optimal too, thanks to a known lower bound
	on the minimum number of names for which
	renaming is solvable wait-free in shared-memory systems,
	whenever $n$ is a power of a prime.
Indeed, our model coincides with the shared-memory model whenever $n=3$,
	and the minimum number of names
	for which renaming is possible in 3-process shared-memory systems is~$5$.
\end{abstract}


\section{Introduction}

\subsection{Motivation}

Two forms of coloring tasks are at the core of distributed computing.
One is \emph{vertex-coloring}~\cite{BarenboimE2013}
	in the framework of synchronous distributed network computing~\cite{peleg2000}.
The other is \emph{renaming}~\cite{AttiyaBDPR90}
	in the framework of asynchronous shared-memory distributed computing~\cite{AttiyaW2004}.
For both tasks, each process starts with its own \emph{identifier} as input,
	which is supposed to be unique in the system,
	and must compute a \emph{color} as output.
The identifiers are supposed to be in a large range of values
(typically of size $\mathrm{poly}(n)$),
	while the colors should lie in a restricted range of values,
	typically $\Set{0,\dots,k-1}$ for some~$k\geq 1$.
Depending on the context,
	$k$ may be an absolute constant,
	or may depend on parameters of the system,
	like the maximum degree~$\Delta$ of the network,
	or even the total number~$n$ of processes.
In the context of network computing,
	the outputs must properly color the underlying graph of the network, i.e.,
	any two neighboring nodes must output distinct colors.
In the context of shared-memory computing,
	each process must output a color that is unique in the system,
	i.e., different from the color of any other process.

On the negative side, it is \enquote{hard}
	to color cycles of even size using only two colors
	in a distributed manner~\cite{Linial92},
	in the sense that $\BigOmega{n}$ synchronous rounds of communication
	are required to solve this problem in the $n$-node cycle~$C_n$
	($n$ even; 2-coloring an odd length cycle is impossible).
A synchronous \emph{round} consists of
	(1)~an exchange of information
	between the two end-points of every edge in the network,
	and (2)~a local computation at every node.
Similarly, renaming the $n$ processes
	of an asynchronous shared-memory system in a wait-free manner
	using a palette with less than $2n-1$ colors
	(i.e., $k$-renaming with $k<2n-1$)
	is impossible~\cite{AttiyaP16,CastanedaR10,HerlihyS99}
	whenever~$n$ is a power of a prime number
	($n=6$ is the smallest integer
	for which this bound does not hold~\cite{CastanedaR12}).
\emph{Wait-free} essentially means that each process
	terminates in a bounded number of write/read steps,
	independently of the asynchronous scheduling of the $n-1$ other processes,
	i.e., independently of the interleaving
	of read and write operations in the shared memory.

On the positive side, it is known that
	3-coloring the $n$-node cycle~$C_n$ for $n\geq 3$
	can be achieved in ${\frac12\log^*n+\BigO{1}}$ synchronous rounds
	thanks to an efficient color-reduction technique
	due to Cole and Vishkin~\cite{ColeV86},
	where $\log^*n$ denotes the number of times
	the $\log_2$ function must be successively applied,
	starting from~$n$, to get a value no greater than~$1$.
This bound is tight, as no algorithms can 3-color the $n$-node cycle in less
than $\frac12\log^*n-1$ rounds, thanks to Linial's celebrated lower bound~\cite{Linial92}.
In shared-memory systems, while $(2n-2)$-renaming
	is impossible wait-free for infinitely many values of~$n$,
	$(2n-1)$-renaming can be achieved wait-free
	for all values of~$n\geqslant 2$~\cite{AttiyaBDPR90}.

The above results are at the core of two separate lines of intensive research.
One line studies extensions of 3-coloring the synchronous cycle,
	in particular $(\Delta+1)$-coloring arbitrary networks
	of maximum degree~$\Delta$;
	see, e.g., \cite{BarenboimEG18,FraigniaudHK16,HalldorssonKMT21,RozhonG20}
	for recent contributions in this field.
This line also studies variants of $(\Delta+1)$-coloring,
	including, for example, $\Delta$-coloring, edge-coloring, weak-coloring, defective coloring;
	see, e.g., \cite{BarenboimE2013,GhaffariHKM21,GhaffariHKMSU20,NaorS95}.
The other line of research studies variants of renaming
	(e.g., long-lived~\cite{AfekAFST99,AttiyaF00}),
	renaming in different shared-memory or message-passing models
	(e.g., \cite{AlistarhAGT12,CastanedaRS12}),
	and the search for algorithms using fewer colors
	whenever $n$ is not a power of a prime~\cite{AttiyaCHP19,CastanedaR12}.

\subsection{Objective}

Our aim is to study coloring tasks
	in a framework relaxing two strong assumptions
	made in the aforementioned contexts.
First, it relaxes the \enquote{all-to-all assumption}
	of the shared-memory model,
	which enables some form of global communication
	between the processing nodes, or processes.
Second, our framework relaxes the \enquote{synchrony assumption}
	of the \local{}~\cite{peleg2000} model of network computing,
	where the processes proceed in lock-step,
	in the sense that we allow \emph{processes}
	to be fully asynchronous and crash-prone,
	while we keep reliable and instantaneous \emph{communications}.
(The latter is in contrast with the classic asynchronous model
	known as \emph{message-passing}~\cite{fischer1985ja},
	where, in addition, the delivery of messages
	is itself asynchronous).
Specifically, we consider a round-based,
	asynchronous computing model in the $n$-node cycle~$C_n$,
	where each \emph{round} of a processing node
	consists of the following sequence of operations:
	(1)~writing in its local register,
	(2)~reading the local registers
	of its two neighbors in~$C_n$, and
	(3)~updating its local state.

The difference with the standard \local{} model,
	in which vertex-coloring is typically studied,
	is that the rounds are asynchronous.
That is, the scheduler may allow some processing nodes
	to perform many rounds
	while other nodes may perform just a few or even zero rounds.
Moreover, the operations performed during a round
	are also asynchronous, e.g.,
	a processing node can write, read,
	and then spend a lot of time idle before changing state.
In particular, the cycle may become disconnected,
	and nodes may become isolated,
	due to nodes that are very slow,
	or even crashed
	(a crash is a full-stop form of failure:
	a crashed node stops functioning,
	and does not recover).
As a consequence,
	information may propagate poorly in the network
	due to slow or crashed nodes.

The difference between our setting and typical models
	(e.g., shared memory)
	used for studying renaming~\cite{AttiyaW2004}
	is that the processing nodes do not share a single array
	of single-writer/multiple-reader registers.
Instead, only adjacent nodes in~$C_n$
	can read each-other's registers.
Thus, instead of having processes
	perform snapshot operations
	-- i.e., read the registers of all processes at once --
	or even immediate snapshots
	-- i.e., write a value \emph{and} read everything all at once --
	each processing node of the cycle
	is restricted to \emph{local} (immediate) snapshots,
	i.e., snapshots that only read the registers
	of its neighboring nodes in the cycle.

We seek to address a few basic questions about this model.
Is wait-free proper vertex-coloring at all possible in~$C_n$?
That is, is renaming at all possible in~$C_n$,
	where renaming in~$C_n$ is the task
	requiring each node to pick a name
	different from the names of its neighbors?
If yes, with how many colors, or, equivalently,
	what is the smallest range of names
	enabling renaming in $C_n$ to be solved?
And what is the minimum number of asynchronous rounds
	to be performed by each node for achieving this task?

Note that it is a priori unclear
	whether wait-free proper vertex-coloring
	is at all possible in the asynchronous cycle,
	even if allowing a large number of colors
	(but less than the number~$n$ of nodes).
Indeed, there are very similar problems
	which are not solvable in this framework.
An example is \emph{maximal independent set} (\MIS{}).
	\textsc{Mis} and 3-coloring are reducible one to another in the cycle
	under the synchronous failure-free setting~\cite{Linial92};
	in contrast, \MIS{} is \emph{not} solvable wait-free
	in the asynchronous crash-prone version of the \local{} model
	considered in this paper.
Indeed, as we detail further down,
a wait-free algorithm for \MIS{}
	could be simulated in the asynchronous shared-memory model
	for solving \emph{strong symmetry-breaking} wait-free,
	which was proved impossible in~\cite{AttiyaP16}.

\subsection{Our Results}

We describe a wait-free algorithm
	for proper coloring the $n$ nodes
	of the asynchronous crash-prone cycle~$C_n$.
So, wait-free proper vertex-coloring is possible in~$C_n$,
	as opposed to, e.g., \MIS{}.
Our algorithm is independent of~$n\geq 3$,
	and each node performs $\BigO{\log^*n}$ asynchronous rounds%
	\footnote{For  every $x>0$, let
	$\log^{(0)}x \Def x$  and, for $k\geq 0$ such that $\log^{(k)}x>0$, let
	$\log^{(k+1)}x \Def \log_2(\log^{(k)}x)$; $\log^*x$ is then defined as the smallest
	$k\geq 0$ such that $\log^{(k)}x \leqslant 1$.} in~$C_n$.
The round-complexity of our algorithm
	is therefore asymptotically optimal
	(up to a constant multiplicative factor),
	thanks to the aforementioned lower bound by Linial~\cite{Linial92},
	which applies even to synchronous (failure-free) executions.

The range of colors used by our algorithm,
	namely $\Set{0,\dots,4}$, is optimal too,
	thanks to the aforementioned minimum number $2n-1$
	of names for which renaming is solvable wait-free in shared-memory systems,
	whenever $n$ is a power of a prime.
Indeed, in the specific case of the cycle~$C_3$,
	our model coincides with the shared-memory model
	with $n=3$ processes,
	which implies that proper coloring~$C_3$
	with less than 5 colors is impossible.

To our knowledge,
	our algorithm is the first distributed coloring algorithm
	designed for a framework combining the following two sources of difficulties:
	on the one hand, asynchrony, and the presence of crash failures
	at any point in time in the computation,
	and, on the other hand, the presence of a network
	limiting the communications between the processing nodes.

\subparagraph{Our Technique.}

Our main algorithm, given in \Cref{alg:fast5},
	has two components. 
	
	The first component of \Cref{alg:fast5} is introduced standalone in \Cref{alg:slow5}
	and bears some resemblance
	to the rank-based $(2n-1)$-renaming algorithm (see \cite[Algorithm
	55]{AttiyaW2004}, and \cite[Step 4 in Algorithm~A]{AttiyaBDPR90}).
It is a wait-free 5-coloring algorithm for~$C_n$,
	i.e., in each of its executions
	over a cycle of length~$n \geqslant 3$, the
	processes that perform enough computational steps
	output a color in~$\Set{0, \ldots, 4}$,
	and no two neighboring processes output the same color.
However, \Cref{alg:slow5} is slow,
	in the sense that its running time may be as large
	as the longest sub-path of the cycle
	along which process identifiers are increasing,
	which can be as large as~$\mathrm{\Theta}(n)$.

The second component of \Cref{alg:fast5}
	uses and modifies the identifiers,
	in parallel to the first component.
This results in quickly shortening sub-paths with increasing identifiers,
	until their length is bounded
	by some constant~$L \leqslant 10$,
	in a manner directly inspired from Cole and Vishkin's
	deterministic coin-tossing algorithm~\cite{ColeV86}.
Each process starts with its input identifiers,
	and successively tries to adopt new ones
	taken from increasingly smaller ranges of identifiers,
	by performing $\BigO{\log^* n}$ identifier-reductions.
As this reduction process goes on,
	the identifiers might not remain unique in the cycle,
	but we ensure that they nonetheless maintain a proper coloring,
	i.e., adjacent processes always hold distinct identifiers.
This invariant is difficult to enforce
	in an asynchronous environment,
	and we resort to a synchronization mechanism
	by which a process awaits a \enquote{green light}
	from both of its neighbors
	each time it aims at changing its identifier.

The second component of our algorithm
	is thus not wait-free by itself,
	since processes are constantly waiting
	for \enquote{green lights} from their neighbors.
However, it offers \emph{starvation free} progress~\cite{herlihy2011pds}.
That is, it is guaranteed to terminate
	when all processes perform
	infinitely many computational steps.
Our core result is that the interaction
	between the wait-free component~1
	and the starvation-free component~2 remains itself wait-free,
	and has a running time $\BigO{\log^* n}$.

We can decompose the description of the first component further,
	in a starvation-free subcomponent
	that looks for a color~$a_p$ for every process~$p$,
	which does not collide with the colors
	of the neighbors of~$p$ with greater identifiers,
	and in another subcomponent that looks
	for a potentially different color~$b_p$ for process~$p$,
	which doesn't collide with the colors of $p$'s neighbors.
That latter subcomponent
	offers \emph{obstruction-free} progress~\cite{herlihy2011pds},
	i.e., the processes terminate whenever they are allowed to take
	multiple consecutive steps by themselves.
As obstruction-free progress and starvation-free progress
	are both strictly weaker than wait-free progress,
	the fact that we bootstrap a wait-free algorithm
	using non-wait-free subcomponents is of independent interest.

\subsection{Related Work}

The recent contributions that are the most related to our work are~\cite{CastanedaDFRR19},
	and the follow-up work~\cite{Delporte-Gallet19}.
The former provides a distributed algorithm for 3-coloring the ring,
	while the latter provides a distributed algorithm
	for $(\Delta+1)$-coloring graphs with maximum degree~$\Delta$.
The model used in these two papers is called \textsf{DECOUPLED},
	as it actually decouples the \emph{computing layer} from the \emph{communication layer}.
Specifically, it assumes $n$ asynchronous crash-prone processes
	occupying the $n$ nodes of a synchronous and reliable network.
That is, the communications remain synchronous,
	and a message emitted by a node~$u$ at round~$r$
	reaches all nodes at distance~$d$ from~$u$ at round~$r+d$.
Moreover, no messages are lost,
	in the sense that a node waking up late
	will still find all messages that passed through it
	stored in a local buffer.
The \textsf{DECOUPLED} model is stronger than
	the fully asynchronous model considered in this paper.
In fact, \cite{Delporte-Gallet19} shows that,
	for every task (e.g., vertex-coloring, edge-coloring, maximal independent set, etc.),
	if there exists a $t$-round algorithm for solving that task in the \textsf{LOCAL} model,
	and if $t=\BigO{\mathrm{polylog}~n}$,
	then there exists a $\BigO{t}$-round algorithm
	for solving the task in the \textsf{DECOUPLED} model.
In contrast, some tasks that are trivial in the \textsf{LOCAL} model --
	and hence in the \textsf{DECOUPLED} model too --
	like 3-coloring $C_3$, or computing a maximal independent set,
	become just impossible in our fully asynchronous model.

The model considered in this paper bears similarities
	with some models used in the context of \emph{self-stabilization}.
Many papers (see, e.g., \cite{BarenboimEG18,BernardDPT09,BlairM12,BlinFB19})
	have addressed the design of self-stabilizing algorithms for 3-coloring the cycles,
	or for $(\Delta+1)$-coloring graphs with maximum degree~$\Delta$.
Self-stabilization assumes that the processing elements
	can behave arbitrarily bad (all variables can be corrupted).
The objective is to design algorithms which,
	starting from an arbitrary initial configuration,
	eventually compute a legal configuration
	(e.g., a configuration in which the colors assigned to the nodes form a proper coloring)
	whenever no failures occur during a sufficiently long period.
In contrast, we assume an initial configuration in which variables are correctly set.
However, we do not assume that the system
	will be failure-free during the execution of the algorithm,
	and the presence of crash-failures
	should not prevent the correct processing nodes from computing a solution.
While 3-coloring the cycle~$C_n$
	is possible in a self-stabilizing manner for all $n\geq 3$,
	$k$-coloring $C_3$ is impossible
	in our fully asynchronous model
	for $k<5$ (\Cref{prop:five-colors}).

\section{Model and Observations}

In this section, we first recall a standard asynchronous variant
	of the (synchronous) \local{} model,
	known as the \emph{state} model.
The model can be viewed
	as a sort of asynchronous message-passing on a graph
	with a local broadcast communication primitive
	and instantaneous message delivery.
Equivalently, it can be viewed
	as a shared-memory system
	where access to the shared memory
	is mediated by a graph;
	we adopt the latter approach in our description.
We define what is a \emph{round} in this model,
	what is the \emph{round complexity} of an algorithm,
	what it means to be \emph{wait-free},
	and then we provide lower bounds on the round-complexity
	and on the range of colors
	for the problem of wait-free vertex-coloring the cycle.

\subsection{Operational Model}

The model is described for the cycle,
	but it can directly be extended to any network.
Specifically, we consider asynchronous wait-free computing
	in the $n$-node cycle~$C_n$,
	where the processes attached to each node
	exchange information between neighbors
	using single-writer/multiple-reader registers.
Each process is a deterministic (infinite) state machine.
All $n$ processes are initially asleep;
	they may wake up at any time,
	and not all processes need to wake up,
	or to take enough steps to terminate correctly
	(i.e., processes are prone to fail-stop faults).
Awakened processes proceed asynchronously,
	each with the objective
	of computing a color in $\Set{0,\dots,4}$.
We focus on wait-free tasks,
	i.e., where a process is guaranteed to terminate
	regardless of the scheduling of the other processes,
	so as to prevent deadlocks
	resulting from a process waiting
	for an event which may never occur
	due to the failure of another process.

Just like for the standard coloring and renaming tasks,
	the only input given to a process~$p$
	is its identifier~$X_p$, which is an integer in the range~%
	$\interval{0}{\mathrm{poly}(n)}$
	that is unique in the system. 
	We do not assume that the processes are aware
	of the length~$n$ of the cycle,
	nor even of an upper bound on~$n$.
Every process proceeds with a sequence
	of exchanges of information with its neighbors
	until some condition is satisfied by its local state,
	at which point it terminates and outputs a color
	obtained by applying some function to this local state.

\subparagraph{Immediate snapshots.}
Let us first recall how communication works
	using a standard \emph{immediate snapshot} communication primitive.
In this model, the~$n$ processes~$p_1, \ldots, p_n$
	communicate through~$n$ single-writer/multiple-readers
	registers~$R_1, \ldots, R_n$,
	initialized with an initial value~$\bot$.
Every process can read all registers,
	but each process~$p_i$ is the single writer in register~$R_i$,
	$i \in \Set{1, \ldots, n}$.
Each process~$p_i$ goes through a (possibly infinite)
	sequence of write-read-update steps,
	where in each step it:
	(1)~writes a value in register~$R_i$,
	(2)~reads the content of all registers, and
	(3)~performs a private computation.
Taken together, these three steps
	constitute an asynchronous \emph{round}
	of process~$p_i$.

Each of the rounds is instantaneous,
	but the time elapsed between two of~$p_i$'s rounds
	may be arbitrarily long.
For example, process~$p_i$ may perform many rounds
	while~$p_j$ performs none,
	in which case~$p_i$ will read the same value
	in register~$R_j$ every time, 
	possibly~$\bot$ if~$p_j$ hasn't awakened yet.
Conversely, in-between two consecutive rounds of~$p_i$'s,
	there may be faster processes
	that performed many writes in their registers.

The value read by a process in a register~$R_j$
	is the one written by~$p_j$ in its most recent round.
Multiple processes may perform a round at the same time. 
	In this case, the system behaves as if
	each of these processes first wrote a value in its own register,
	then all processes read all registers,
	and, finally, they all performed their private computation.
Note that distinct processes
	may be at distinct rounds of their execution.
For example, one process may be just starting,
	i.e., in its first round,
	while another may already have been running for some time,
	and so be at a later round.

\subparagraph{Local immediate shapshots.}

Our model simply adds a graph to the above,
	which mediates which registers
	a process is able to read.
For example, in the cycle,
	a process only reads three registers:
	its own register, and the register of each of its two neighbors.
We do not assume a coherent notion of left and right,
	i.e., each node assigns an arbitrary order
	to the registers of its neighbors.

In this paper, we do not assume that the registers are bounded. Nevertheless, our algorithms only manipulate
	a constant number of variables using~$\BigO{\log n}$ bits each.

\subsection{Schedules and complexity}

In our model, an execution is entirely characterized
	by the code of each process, the graph (here, the cycle~$C_n$),
	the input identifiers of each process,
	as well as the activation patterns of each process.
The latter is captured by the collection
	of~$n$ increasing sequences~$t_p^{(1)}, t_p^{(2)}, \ldots$
	of positive integers, one for each process~$p \in [n]$,
	where~$t^{(i)}$ denotes the time in which
	process~$p$ performs its~$i$-th round.

As multiple processes may be performing rounds simultaneously,
	let us introduce, for~$t \geqslant 1$,
	the set $\sigma(t)$ of activated processes at time~$t$.
We set: $p \in \sigma(t) \iff \exists i \geqslant 1 : t_p^{(i)} = t$.
The ""schedule"" of an execution
	is the infinite sequence~%
	$\sigma = \sigma(1), \sigma(2), \ldots$
	An execution of a given algorithm on the cycle~$C_n$
	is thus determined by the schedule~$\sigma$
	and the input identifiers~$(X_p)_{p \in [n]}$.

We will say that a process~$p \in \sigma(t)$ is ""working""
	if the stopping condition of~$p$
	has not been fulfilled before time~$t$.
This leads us to define, for any schedule~$\sigma$,
	the restricted schedule~$\overline{\sigma}$ of working processes:
\[
	\overline{\sigma}(t) \Def
	\Set{p \in \sigma(t) \Where
	\text{$p$ has not fulfilled the stopping condition
	at time $\leq t-1$}}.
\]
An execution ""terminates"" if there exists some time~$t^*$
	such that~$\overline{\sigma}(t) = \emptyset$
	for all subsequent times~$t \geqslant t^*$,
	i.e., if eventually all processes stop working.
	Note that a process stops working according to two possible scenarios: it may have been activated sufficiently many times for allowing it to fulfill the stopping condition, or it was not activated after some time~$t$, before it fulfilled the stopping condition. The latter scenario models the crash of a process (at time $t$, or earlier in the execution). 
The ""round complexity"" of a terminating execution
	is then defined as 
	\[
	\max \Set{i \in \Natural \Where \exists p \in [n] : p \in \overline{\sigma}(t_p^{(i)})}.
	\]
The running time of an algorithm over the cycle~$C_n$
	is then the supremum of the round complexity
	\emph{for all possible executions}, i.e.,
	all possible identifier assignments and schedules. 
	Informally, the running time corresponds to
	the maximal number of times a process can be activated
	before it is guaranteed to terminate.
An algorithm is then ""wait-free""
	if its running time is finite.

\subsection{Lower Bounds and Impossibility Results}

We complete this section by a couple of observations
	on the round-complexity,
	and on the range of colors used
	by wait-free vertex-coloring algorithms for the cycle.
Before that, we formalize the fact that,
	as claimed in the introduction,
	the maximal independent set (\MIS{}) problem
	cannot be solved in the asynchronous cycle.
	Solving the \MIS{} problem requires that,
	at the end of every execution,
	(1)~every node that terminates and outputs~0
	is neighbor of at least one node that terminates and output~1,
	and (2)~no two neighboring nodes that terminate output~1.

\begin{property}
For every $n\geq 3$,
\MIS{} in the $n$-node cycle~$C_n$,
	cannot be solved wait-free in our model.
\end{property}

\begin{proof}
The proof is by reduction from the \emph{strong symmetry-breaking} (SSB) problem,
	which cannot be solved wait-free
	in the asynchronous shared-memory model
	(see~\cite[Theorem 11]{AttiyaP16}).
We show that if there were an algorithm
solving \MIS{} in the $n$-node cycle,
	then there would exist an algorithm for SSB
	in the $n$-node shared-memory system.
Recall that SSB requires that
	(1)~if all processes terminate,
	then at least one processes outputs~0,
	and at least one process outputs~1,
	and (2)~in every execution,
	at least one process outputs~1.
By way of contradiction,
let $A$ be an algorithm solving \MIS{} in $C_n$.
The $n$~processes of shared-memory system can simulate $A$ as follows.
Process $p_i$, $i=0,\dots,n-1$,
	simulates the execution of~$A$ at the node of $C_n$ with identifier~$i$,
	and with neighbors the nodes with identifiers $i\pm 1\bmod n$,
	which are simulated by processes $p_{i\pm 1\bmod n}$, respectively.
	Since $A$ solves \MIS{},
	it guarantees that, if all processes terminate,
	then at least one outputs~0,
	and at least one outputs~1.
Moreover, in every execution of~$A$,
	a node that terminates and is isolated
	(none of its neighbors terminated) must output~1,
	and a node that terminates and has a neighbor that terminates
	is such that either itself outputs~1,
	or at least one of its neighbors outputs~1.
This guarantees that, in every execution,
	at least one process output~1.
The two conditions for solving SSB
	are therefore fulfilled by simulating~$A$,
	and thus $A$ cannot exist.
\end{proof}

We now show that the round-complexity
	of our vertex-coloring algorithm is optimal.

\begin{property}
For every $k\geq 2$,
	the round-complexity of any wait-free algorithm
	for $k$-coloring the vertices
	of the $n$-node cycles~$C_n$, $n\geq 3$,
	requires $\BigOmega{\log^*n}$ rounds in our model.
\end{property}

\begin{proof}
This directly follows from~\cite{Linial92},
	which proved that, even in the synchronous failure-free setting,
	$k$-coloring the vertices of the $n$-node cycles~$C_n$,
	requires $\BigOmega{\log^*n}$ rounds.
\end{proof}

Finally, we show that the range of colors
	used by our vertex-coloring algorithm
	is optimal over the class of all cycles.

\begin{property}\label{prop:five-colors}
In our model, any wait-free $k$-coloring algorithm for the $n$-node cycles~$C_n$, $n\geq 3$, satisfies $k\geq 5$.
\end{property}

\begin{proof}
For a cycle of length~$n=3$,
	our model coincides with
	the standard wait-free shared-memory model with immediate snapshots.
The result thus directly follows
	from the fact that renaming among $n=3$ processes
	cannot be done using fewer than 5~names,
	in shared-memory systems with immediate snapshots~\cite{AttiyaP16,CastanedaR10}.
\end{proof}

Note that the above property does not imply that,
	for specific values of~$n$,
	fewer colors could not be used,
	the same way the lower bound $2n-1$
	on the number of names for renaming
	holds for $n$~power of a prime only.
However, a generic algorithm capable of proper coloring
	all cycles $C_n$, for all $n\geq 3$,
	must use at least 5~colors,
	as our algorithm does.
Nevertheless, for $n>3$, the state model
	and the shared-memory model with immediate snapshots
	do not coincide,
	and thus it may well be the case that fewer than~5 colors
	could be used for some specific values of~$n>3$,
	although we conjecture that this is not the case.

\section{Asynchronously coloring the cycle in linear time}\label{section:slow5}

Here we develop asynchronous coloring algorithms
	that we show to guarantee wait-free progress
	-- i.e., a process will terminate in all executions,
	provided that it is activated sufficiently many times --
	and correct -- i.e., the graph induced by the terminating processes
	is properly colored by the output colors of these processes.
These algorithms have a poor runtime complexity of~$\BigO{n}$ steps
	when compared to state-of-the-art algorithms in the \local{} model,
	which terminate in~$\BigO{\log^* n}$ synchronous rounds.
We will achieve a similar runtime complexity in the next section
	by augmenting our wait-free algorithms with mechanisms speeding up termination.

We first present an algorithm that uses a~6-color palette.
Although it uses one extra color
	when compared to the theoretical minimum of~$5$ colors
	required to color (i.e., rename) the cycle~$C_3$,
	this allows us to illustrate some of our main algorithmic ingredients.
We then present another wait-free algorithm
	that colors any cycle using a~5-color palette.

\subsection{Warmup: using a palette of six colors}

In \Cref{alg:slow6}, we present a simple algorithm
	for wait-free coloring any cycle~$C_n$ ($n \geq 3$),
	using the six colors in the set~$\Set{(a,b) \in \Natural \times \Natural \Where a + b \leqslant 2}$.
Given a process $p$, we denote by $X_p$ its identifier,
	and by $q$ and $q'$ its two neighboring process in the cycle.
We denote by $u \sim v$ the fact
	that processes $u$ and $v$ are neighbors in~$C_n$.
A process~$p$, with neighbors~$q$ and~$q'$,
	is said to be ""locally extremal""
	(with respect to the identifiers)
	if either~$X_p > \max \Set{X_q, X_{q'}}$
	or~$X_p < \min \Set{X_q, X_{q'}}$.

Intuitively, \Cref{alg:slow6} guarantees that
	locally extremal processes quickly terminate,
	by sticking to one of the two components~$a_p$ or~$b_p$
	of their color~$c_p = (a_p,b_p)$
	(\Cref{lemma:slow6:stubborn}).
Termination then propagates throughout the cycle,
	due to the wait-free nature of the algorithm
	(\Cref{lemma:slow6:obfree,lemma:slow6:stubborn}).
Given an initial coloring of~$C_n$
	provided by the nodes' identifiers,
	we will show that the worst-case convergence time of a process
	is determined by its distance
	to its nearest local extrema,
	which is bounded by~$\BigO{\min\Set{n, \max_p X_p - \min_q X_q}}$,
	which yields a linear convergence time.

\begin{algorithm}
	\caption{6-coloring algorithm, code for process~$p$ with neighbors~$q$ and~$q'$}\label{alg:slow6}
	\Input{$X_p \in \Natural$}
	\Initially{
		$c_p = (a_p, b_p) \gets (0,0) \in \Natural \times \Natural$
	}
	\Forever{
		\textbf{write}$(X_p, c_p)$ \text{and} \textbf{read}$((X_q, c_q),(X_{q'}, c_{q'}))$ \Comment{local immediate snapshot}
		\lIf{$c_p \notin \Set{c_q,c_{q'}}$}{\Return{$c_p$}}\label{alg:slow6:return}
		\Else{
			$a_p \gets \min \Natural
			\setminus \Set{a_u \Where (u \sim p) \wedge (X_u > X_p)}$
			\label{alg:slow6:alpha}\;
			$b_p \gets \min \Natural
			\setminus \Set{b_u \Where (u \sim p) \wedge (X_u < X_p)}$
			\label{alg:slow6:beta}\;
		}
	}
\end{algorithm}

\begin{theorem}\label{theo:alg:slow6}
	In any execution of \Cref{alg:slow6} over the cycle~$C_n$
		with a proper coloring provided by the values $(X_p)_{p\in [n]}$
		given to the processes as input, we have:
		\begin{description}
			\item[\textsf{Termination:}]
				every process terminates
					after having been activated
					at most $\Floor{3n/2} + 4$ times;
			\item[\textsf{6-color palette:}]
				every process that terminates outputs a color
				in the set~$\Set{(a,b) \Where a + b \leqslant 2}$;
			\item[\textsf{Correctness:}]
				the outputs properly color the graph
				induced by the terminating processes in~$C_n$.
		\end{description}
\end{theorem}

The rest of the subsection is dedicated to the proof of \Cref{theo:alg:slow6}.
Recall that, in a schedule $\sigma$,
	a process~$p \in \sigma(t)$ is ""working"" in~$t$
	if it has not returned before~$t$.
We will say of a process
	that "works" at time~$t$ and terminates at that time
	that it ""succeeds"",
	and otherwise that it ""misses"".
A process may "succeed" only once,
	after which it no longer partakes in the execution.

\subparagraph{Notation.}

We will adopt the following notation
	for all algorithms throughout the paper.
	If $x_p$ is a variable used by process $p$,
	we use $x_p(t)$ to denote
	the value of $x_p$ \emph{in $p$'s memory},
	at the \emph{end} of time~$t$,
	and we use $\hat{x}_p(t)$ to denote the value of $x_p$
	\emph{visible to $p$'s neighbors}
	at the end of time $t$.
Let $x_p(0)$ be given by the initialization of the algorithm,
	and let $\hat{x}_p(0) = \bot$.
By definition, we have
	\begin{equation}\label{eq:hat}
		\hat{x}_p(t) =
		\begin{cases}
			x_p(t-1) & p \in \overline{\sigma}(t) \\
			\hat{x}_p(t-1) & p \notin \overline{\sigma}(t)
		\end{cases}
	\end{equation}

\begin{lemma}\label{lemma:slow6:avoid}
	Let $t \geq 0$, and let $p \in \overline{\sigma}(t)$.
	We have $c_p(t) \notin \Set{\hat{c}_q(t) \Where q \sim p}$,
		and process~$p$ returns at time~$t$
		if and only if~$c_p(t) = c_p(t-1)$.
\end{lemma}

\begin{proof}
	Process~$p$ does not update~$c_p$ when it returns,
		and so $c_p(t) = c_p(t-1)$ whenever $p$ returns at time~$t$.
	Let us then assume that~$p \in \overline{\sigma}(t)$
		does \emph{not} return at time~$t$,
		and let~$q$ be one of~$p$'s neighbors.
	If~$q$ has not yet been activated then
		$\hat{c}_q(t) = \bot \neq \hat{c}_p(t)$.
	If $q$ has been already activated then, since the inputs form an initial proper coloring,
		we either have~$X_p > X_q$ or $X_p < X_q$.
	In the former case, we have~$a_p(t) \neq \hat{a}_q(t)$,
		and in the latter case, we have~$b_p(t) \neq \hat{b}_q(t)$.
	Either way, we have~$c_p(t) \neq \hat{c}_q(t)$,
		and so~$c_p(t) \neq c_p(t-1)$, since~%
		$c_p(t-1) = \hat{c}_p(t)
		\in \Set{\hat{c}_q(t),\hat{c}_{q'}(t)}$
\end{proof}

Lemma~\ref{lemma:slow6:avoid} provides us
	with an effective characterization of $\overline{\sigma}$:
	for every $t\geq 0$ and every~$p\in[n]$,
	\begin{equation}
		p \in \overline{\sigma}(t) \iff \forall t' < t  :
			\big(p \in \sigma(t') \implies c_p(t') \neq c_p(t'-1)\big).
	\end{equation}
The next lemma formalize the intuition 
	that a process terminates fast,
	unless the execution is \enquote{very interleaved}.

\begin{lemma}\label{lemma:slow6:obfree}
	Let $p$ be a process
		that is working at times $t_1$ and $t_2>t_1$,
		but is not activated
		at any time~$t \in \interval{t_1+1}{t_2}$.
	If neither of $p$'s neighbors
		is working in the time interval~$(t_1,t_2)$,
		then process~$p$ returns at time~$t_2$.
\end{lemma}

\begin{proof}
	The result directly follows from Lemma~\ref{lemma:slow6:avoid}, using the fact that~$c_p(t_1) \notin \Set{\hat{c}_q(t_1) \mid q \sim p}$
		and ${\hat{c}_p(t_2) = c_p(t_1)}$.
\end{proof}

As the next lemma shows,
	a process cannot be prevented from returning
	by only one of its neighbors.

\begin{lemma}\label{lemma:slow6:stubborn}
	Let process~$p$ be activated
		at times~$t_1 < t_2 < t_3 < t_4$,
		but not at any other time $t \in (t_1,t_4)$. 
		If $a_p(t_1) = a_p(t_2) = a_p(t_3) = a_p(t_4)$,
		and $X_p$ is not a local minimum,
		then $p$ returns at time at most~$t_4$.
	The same holds if~$b_p(t_1) = b_p(t_2) = b_p(t_3) = b_p(t_4)$
		and~$X_p$ is not a local maximum.
\end{lemma}

\begin{proof}[Proof of~\Cref{lemma:slow6:stubborn}]
	We establish the result
		for the case where~$X_p$ is not a local minimum
		and $a_p(t_1) = a_p(t_2) = a_p(t_3) = a_p(t_4)$.
	The proof uses the same arguments
		with local maxima
		and $b_p(t_1) = b_p(t_2) = b_p(t_3) = b_p(t_4)$.
	
	Suppose that process~$p$ fails to return at time~$t_1$;
		we consider two cases.
		
	If~$p$ is a local maximum,
		then we have~$\hat{a}_p(t) = 0$ for all~$t$.
	Moreover,
		if some process~$q \sim p$
		is working in the interval~$\interval{t_1}{t_3}$,
		then~$a_q(t_3) \neq 0$.
	Furthermore, we have $c_p(t_3) \neq \hat{c}_q(t_3)$ by \Cref{lemma:slow6:avoid}.
	In this case, either process~$q$ works
		in the interval~$\interval{t_3+1}{t_4}$,
		and~$\hat{a}_q(t_4) \neq 0$, 
		or it does not work in this interval,
		and~$\hat{c}_q(t_4) = \hat{c_q}(t_3)$.
	Either way, we get $\hat{c}_p(t_4) \neq \hat{c}_q(t_4)$,
		and thus $p$ returns at time at most~$t_4$.
	Suppose now process~$p$ has a neighbor $q'$
		that is inactive in the interval~$\interval{t_1}{t_3}$.
	If~$p$'s other neighbor~$q$ is \emph{not} working
		in the interval~$\interval{t_1+1}{t_2}$,
		then~$p$ returns at time~$t_2$
		by \Cref{lemma:slow6:obfree};
		if, on the other hand,
		$q$ is working in this interval,
		then we have~$a_q(t_2) \neq 0$,
		and, as above, process~$p$ returns at time at most~$t_3$.

	If process~$p$ is \emph{not} a local maximum,
		then it has a neighbor~$q'$ with~$X_{q'} > X_p$.
	If we had $\hat{a}_p(t)=\hat{a}_{q'}(t)$,
		in any~$t \in \Set{t_2, t_3, t_4}$,
		then ${a}_p(t)$ would be switched,
		which contradicts the lemma assumptions;
		hence we have~$\hat{c}_p(t) \neq \hat{c}_{q'}(t)$
		for~$t = t_2, t_3, t_4$.
	Suppose~$p$ fails to return in~$t_2$.
	In this case, as before, either the other neighbor~$q$ of~$p$
		is working in the interval~$\interval{t_2+1}{t_3}$,
		and so~$a_q(t_3) \neq \hat{a}_p(t_4)$;
		or~$q$ is inactive in that interval
		and~$p$ returns at time~$t_3$.
	Either way, process~$p$ returns
		at the latest at time~$t_4$. 
\end{proof}

Note that, even though $X_p(t)$ remains constant throughout the execution,
	the public value $\hat{X}_p(t)$ doesn't,
	as initially its value is~$\bot$.
	To analyse executions of \Cref{alg:slow6},
	let us introduce the sets
	\begin{equation*}
	\Ngt_p(t) \Def \Set{q \sim p \Where  \hat{X}_q(t) > \hat{X}_p(t)} \;
	\text{and}\; \Nlt_p(t) \Def \Set{q \sim p \Where \hat{X}_q(t) < \hat{X}_p(t)}.
	\end{equation*}
We furthermore define the sets
	\begin{equation}\label{eq:setsAB}
		A_p(t) \Def
		\begin{cases}
			\bigcup_{q \in \Ngt_p(t)} \Brackets*{\hat{A}_q(t) \cup \Set{\hat{X}_q(t)}}
			& p \in \overline{\sigma}(t) \\
			A_p(t-1) & p \notin \overline{\sigma}(t)
		\end{cases}
	\end{equation}
and
	\begin{equation}\label{eq:setsABbis}
			B_p(t) \Def
		\begin{cases}
			\bigcup_{q \in \Nlt_p(t)} \Brackets*{\hat{B}_q(t) \cup \Set{\hat{X}_q(t)}}
			& p \in \overline{\sigma}(t) \\
			B_p(t-1) & p \notin \overline{\sigma}(t)
		\end{cases}
	\end{equation}
	where~$A_p(0) = B_p(0) = \varnothing$,
	and where the sets~$\hat{A}_p(t), \hat{B}_p(t)$
	are defined according to \Cref{eq:hat}.
The set~$A_p(t)$ contains
	all processes that~$p$ has heard of at time~$t$,
	and that are linked to~$p$ through a subpath of~$C_n$
	where process identifiers are increasing.
Symmetrically, the set~$B_p(t)$ contains processes
	that~$p$ has heard of, and that are linked to~$p$
	through a subpath where identifiers are decreasing.

\begin{lemma}\label{lemma:setsABexclude}
	Let $t \in \Natural$, and let $p\in[n]$ be a process. For every $x\in A_p(t)$, we have $\hat{X}_p(t) < x$, and, for every $x\in B_p(t)$, we have $\hat{X}_p(t) > x$. 
\end{lemma}

\begin{proof}[Proof of~\Cref{lemma:setsABexclude}]
	We proceed by induction on~$t \in \Natural$.
	For~$t = 0$, the claim is vacuously true,
		as~$A_p(t) = B_p(t) = \varnothing$.
	For the inductive step, we suppose the claim holds for~$t = 0, \ldots, T$,
		and we  show that it holds for~$t = T+1$.
	If~$p \notin \overline{\sigma}(T+1)$ then
		we have~$\hat{X}_p(T+1) = \hat{X}_p(T)$ and~$A_p(T+1) = A_p(T)$,
		Thus the claim holds by induction.
	Let us assume that~$p \in \overline{\sigma}(T+1)$, and let~$x \in A_p(t)$.
	By \Cref{eq:setsAB} and the assumption~$p \in \overline{\sigma}(T+1)$,
		there exists~$q \in \Ngt_p(T+1)$
		such that either~$x = \hat{X}_q(T+1)$ or~$x \in \hat{A}_q(T+1)$.
	In the former case, we have~$\hat{X}_p(T+1) < Y$ by the definition of~$\Ngt_p$.
	In the latter case, there must exist some time~$t' \leqslant T$,
		with~$q \in \overline{\sigma}(t')$,
		for which~$\hat{X}_q(T+1) = X_q(t')$ and~$\hat{A}_q(T+1) = A_q(t')$.
	Since~$\tau \leqslant T$,
		we get that that $\hat{X}_q(t') < x$,
		thanks to the induction hypothesis.
	Also, since the value of~$X_q(t)$
		is stable throughout the execution,
		we have~$X_q(t') = X_q(t'-1) = \hat{X}_q(t') < x$.
	Therefore~$\hat{X}_q(T+1) < x$,
		and, since~$q \in \Ngt_p(T+1)$, 
		we have~$\hat{X}_p(T+1) < \hat{X}_q(T+1) < x$,
		which proves the claim.

	The proof is symmetric for~$x \in B_p(t)$.
\end{proof}

\begin{remark}\label{remark:size}
	This will be used in the next section,
	where we present a procedure for
	speeding up \Cref{alg:slow6} by reducing the space of colors
	initially provided to the nodes thanks to their identifiers.
	On the other hand, the claim~$\hat{X}_p(t) > \max B_p(t)$
		doesn't generalize under the same weaker condition.

In the case where $X_p$ does not change, we can notice that $A_p(t)$ and $B_p(t)$ are increasing, inclusion-wise, with time.
Moreover, the elements of~$A_p(t)$
	correspond to increasing identifiers~$X_q$
	following a path from $p$
	(decreasing in the case of $B_p(t)$).
Hence, $\abs{A_p(t)}$ has a size
	bounded by the length of
	the longest path of increasing identifiers from $p$.
\end{remark}

If a process~$p \in \overline{\sigma}(t)$
	fails to return in time~$t$,
	the sets~$A_p(t)$ and~$B_p(t)$
	help us compute its next color~$c_p(t)$.

\begin{lemma}\label{lemma:slow6:parity}
	For any time~$t \geqslant 1$,
		if a process~$p \in \overline{\sigma}(t)$
		fails to return at time~$t$, then:
		\begin{enumerate}
			\item if $\abs*{\Ngt_p(t)} \leqslant 1$,
				then $a_p(t) \equiv \abs*{A_p(t)} \bmod 2$;
			\item if $\abs*{\Nlt_p(t)} \leqslant 1$,
				then $b_p(t) \equiv \abs*{B_p(t)} \bmod 2$.
		\end{enumerate}
\end{lemma}

\begin{proof}[Proof of~\Cref{lemma:slow6:parity}]
	We only treat the case~$\abs*{\Ngt_p(t)} \leqslant 1$,
		as the other case is symmetric.
First, note that for any process~$q$,
		$\hat{X}_q(t)$ is equal to either~$\bot$ or~$X_q$.
	In the former case,
		process~$q$ is still inactive in time~$t$,
		and thus~$A_q(t) = \varnothing$.
	As a consequence, thanks to \Cref{lemma:setsABexclude},
		we have~$X_q \notin A_q(t)$ for all~$t \in \Natural$.

	Given~$p \in \overline{\sigma}(t)$,
		we proceed by induction over~$\abs*{A_p(t)}$
		by treating two base cases~$\abs*{A_p(t)} = 0$, $\abs*{A_p(t)} = 1$,
		and then the general case. For the base cases, as~$p \in \overline{\sigma}(t)$,
		we have  $\abs*{A_p(t)} = 0$ if and only if $\Ngt_p(t) = \emptyset$,
		which corresponds to~$p$ being a local maximum
		among its neighbors awaken at time~$t$.
	In this case, if~$p$ fails to return,
		then the algorithm enforces~$a_p(t) = 0$, as desired,
		which gives the base case of the induction.
	If $\abs*{A_p(t)} = 1$,
		then the set~$\Ngt_p(t)$ is a singleton. Let $\Set{q}=\Ngt_p(t)$. 
	We have $A_p(t) = \Set{\hat{X}_q(t)} = \Set{X_q}$,
		and~$\hat{A}_q(t) \in \Set{ \varnothing, \Set{X_q}}$.
	The set~$\hat{A}_q(t)$ is therefore empty, i.e., 
		$\hat{a}_q(t) = 0$,
		and thus the algorithm enforces $a_p(t) = 1 = \abs*{A_p(t)}$.

	For the inductive case, let us assume that the claim is true
		for~$\abs*{A_p(t)} = 0, \ldots, T$ with $T \geqslant 1$,
		and let us show that it still holds
		for~$\abs*{A_p(t)} = T+1 \geqslant 2$.
	Here again, the set~$\Ngt_p$ has to be a singleton, say~$\Set{q}$,
		and so we have~$a_p(t) = 1 - \hat{a}_q(t)$,
		and~$A_p(t) = \Set{X_q} \cup \hat{A}_q(t)$,
		with $X_q \notin \hat{A}_q(t)$.
	Thus~$\abs*{\hat{A}_q(t)} = T$,
		and there was an earlier time~$t' < t$
		where~$q \in \overline{\sigma}(t')$ failed to return, and~$\hat{A}_q(t) = A_q(t')$.
	Since~$X_p < X_q$,
		$\abs{\Ngt_q(t')} \neq 2$,
		and so $a_q(t') \equiv T \pmod 2$ by the induction hypothesis. 
		Thus~$a_p(t) = 1 - a_q(t') \equiv T+1 \pmod 2$,
		which completes the proof of the claim.
\end{proof}

As a direct consequence of Lemma~\ref{lemma:slow6:parity},
	we get the following.

\begin{lemma}\label{lemma:slow6:2sides}
	Let $t\geq 0$, and let $p \in [n]$
		be non-extremal a process.
	If~$p \in \overline{\sigma}(t)$,
		but $p$~fails to return at time~$t$,
		then we have $A_p(t) \neq A_p(t-1)$ or~$B_p(t) \neq B_p(t-1)$.
\end{lemma}

\begin{proof}
	 Using \Cref{lemma:slow6:parity},
		if~$A_p(t) = A_p(t-1)$ and~$B_p(t) = B_p(t-1)$
		then~$c_p(t) = c_p(t-1)$,
		and so by \Cref{lemma:slow6:avoid} process~$p$ returns,
		a contradiction.
\end{proof}

This leads us to the following complexity bound
	for processes that are not local extrema.
It relies on the distance of a process
	to its closest local extrema along monotone paths.
Let $q_i$, $i=0,\dots,k+1$,
	be a set of distinct processes,
	excepted possibly $q_{k+1} = q_0$.
Let us assume that these processes
	form a subpath of~$C_n$,
	or possibly the entire cycle~$C_n$ if $q_{k+1} = q_0$.
That is, $q_0 \sim q_1 \sim q_2 \cdots \sim q_k \sim q_{k+1}$.
Let us assume that $X_{q_0} < X_{q_1}$ and $X_{q_k} < X_{q_{k+1}}$, but
$X_{q_1} > X_{q_2} > \cdots > X_{q_k}$,
	i.e., process~$q_1$ is locally maximal,
	process~$q_k$ is locally maximal, and
	for $i\in\Set{1,\dots,k}$,
	process~$q_i$ is at monotone distance~$i-1$
	from its closest local maximum $q_1$,
	and at monotone distance~$k-i$
	from its closest local minimum~$q_k$.

\begin{lemma}\label{lemma:slow6:complexity}
	Let $p\in[n]$ be a non-extremal process,
		and let $\ell$ and~$\ell'$ be the monotone distances
		from~$p$ to its closest extremal processes.
	Process~$p$ returns after at most~$\min\Set{3\ell, 3\ell', \ell + \ell'} + 4$ activations.
\end{lemma}

\begin{proof}
We know from \Cref{remark:size} that $A_p(t)$ is increasing with time, and that its size is bounded by $\ell$. Thanks to  \Cref{lemma:slow6:parity}, we have that $a_p(t)$ is determined by the size of $A_p(t)$. It follows that $a_p(t)$ changes at most $\ell+1$ times. Symmetrically, $b_p(t)$ changes at most $\ell'$ times.
By \Cref{lemma:slow6:stubborn},
	we get that a process $p$
	cannot be activated more than 3 times
	while keeping the same value for~$a_p(t)$.
It follows that process~$p$ can be activated
	at most $3\ell+4$ times before it returns.
Symmetrically, $p$ can be activated
	at most $3\ell'+4$ times before it returns.
Finally, from \Cref{lemma:slow6:2sides},
	we get that $p$ can be activated
	at most $\ell+\ell'+1$ times before it returns.
\end{proof}

This last result allows us to conclude:

\begin{proof}[Proof of \Cref{theo:alg:slow6}]
As a direct corollary of \Cref{lemma:slow6:stubborn},
	that local extrema return after at most $4$ steps:
	a maximum will maintain~$a(t)=0$, and a minimum, $b(t)=0$.
For the other nodes,
	\Cref{lemma:slow6:complexity} gives us the complexity,
	knowing that $\min\Set{\ell,\ell'}$ is bounded by $\Floor{3n/2}$.
\end{proof}

\begin{remark}
	\Cref{lemma:slow6:complexity} states that
		the complexity of \Cref{alg:slow6} is linear
		in the length of the longest chain of processes
		$p_1 \sim p_2 \sim \cdots$
		that is monotone for the identifiers, i.e.,
		$X_{p_1} > X_{p_2} > \cdots$.
	Throughout this section,
		we have assumed that the processes start with their identifiers as input, and that each identifier is unique in the network,
		i.e., $X_p \neq X_q$ whenever~$p \neq q$.
	Note however
		that \Cref{theo:alg:slow6} only requires that identifiers form a \emph{proper coloring},
		i.e., $X_p \neq X_q$ whenever~$p \sim q$.
	In this case, the length of a monotone chain
		is bounded by the number of initial colors,
		and so is the convergence of \Cref{alg:slow6}.
	In the \Cref{section:fast5},
		we exploit this property
		to dramatically accelerate \Cref{alg:slow6}
		by dynamically adjusting the \enquote{identifiers}~$X_p$ themselves,
		using a modification
		of Cole and Vishkin's classic algorithm~\cite{ColeV86},
		initially designed for the \textsf{PRAM} model,
		but easily adapted to the \textsf{LOCAL} model.
	As we shall see, its adaptation
		to the asynchronous setting is more subtle.
\end{remark}

\subsection{Saving one color: wait-free 5-coloring the cycle}

Here we present, in \Cref{alg:slow5},
	another wait-free coloring algorithm for the cycle,
	which only uses a palette of five colors.
As already noted, when the graph is a clique,
	asynchronous coloring is identical to the renaming problem
	using an immediate snapshot communication primitive,
	which implies that asynchronously coloring the cycle~$C_3$
	\emph{requires} at least a 5-color palette.
Our algorithm is thus optimal in terms of colors
	for the class ${\mathcal{C} = \Set{C_n \Where n \geqslant 3}}$ of all cycles. 

\begin{algorithm}
	\caption{5-coloring algorithm, code for process~$p$ with neighbors $q$ and $q'$}\label{alg:slow5}
	\Input{$X_p \in \Natural$}
	\Initially{
		$a_p, b_p \gets 0 \in \Natural$\;
	}
	\Forever{
		\textbf{write}$(X_p, a_p, b_p)$
		\text{and} \textbf{read}$((X_q, a_q, b_q),(X_{q'}, a_{q'}, b_{q'}))$
		\Comment{local imm. snap.}
		$P^{+} \gets \Set{u \in \Set{q,q'} \Where X_u > X_p}$\;
		$C^{+} \gets
			\Set{a_u \Where u \in P^{+}} \cup
			\Set{b_u \Where u \in P^{+}}$
			\label{alg:slow5:ceeplus}\;
		$C \gets \Set{a_q, b_q, a_{q'}, b_{q'}}$\label{alg:slow5:cee}\;
		\lIf{$a_p \notin C$}{\Return{$a_p$}}
		\lElseIf{$b_p \notin C$}{\Return{$b_p$}}
		\Else{
			$a_p \gets \min \Natural \setminus C^{+}$\;
			$b_p \gets \min \Natural \setminus C$\;
		}
	}
\end{algorithm}

\begin{theorem}\label{theo:alg:slow5}
	In any execution of \Cref{alg:slow5} over the cycle~$C_n$
		with a proper coloring
		provided by the values $(X_p)_{p\in [n]}$
		given to the processes as input, we have:
		\begin{description}
			\item[\textsf{Termination:}]
				every process terminates
					after having been activated
					at most~$\BigO{n}$ times;
			\item[\textsf{5-color palette:}]
				every process that terminates
				outputs a color in the set~$\Set{0, \ldots, 4}$;
			\item[\textsf{Correctness:}]
				the outputs properly color the graph
				induced by the terminating processes in~$C_n$.
		\end{description}
\end{theorem}

From the algorithm,
	we immediately deduce the following characterization
	of when a process returns a value.

\begin{lemma}\label{lemma:slow5:avoid}
	Let $t \geq 1$, and let $p \in \overline{\sigma}(t)$
		be a process with neighbors~$q$ and~$q'$. 
		Let $C \coloneqq \Set{\hat{a}_q(t), \hat{b}_q(t), \hat{a}_{q'}(t),
		\hat{b}_{q'}(t)}$.
	We have~$b_p(t) \notin C$,
		and process~$p$ returns at time~$t$
		if and only if~$a_p(t-1) \notin C$ or~$b_p(t-1) \notin C$.
\end{lemma}

Note that, as a consequence of the previous lemma, 
	$b_p(t) \neq b_p(t-1)$
	unless~$p \in \overline{\sigma}(t)$
	returns at time~$t$.
Thus \Cref{lemma:slow6:obfree}
	continues to hold for \Cref{alg:slow5}.

Defining the sets~$A_p(t)$
	as we did for \Cref{alg:slow6},
	we get the following sufficient condition
	for a process to terminate.

	\begin{lemma}\label{lemma:slow5:Ap}
	Suppose that process~$p \in [n]$
		is \emph{not} a local minimum for the identifiers.
	If~$p$ is activated at times~$t_1 < t_2 < t_3 < t_4$,
		and~$A_p(t_1) = A_p(t_2) = A_p(t_3) = A_p(t_4)$,
		then~$p$ returns at time at most~$t_4$.
\end{lemma}

\begin{proof}[Proof of~\Cref{lemma:slow5:Ap}]
	We first show the following:
		if~$p \in \overline{\sigma}(t)$
		fails to return at time~$t \geqslant 1$, then
		\begin{equation}
			a_p(t) = 0 \iff \abs*{A_p(t)} \equiv 0 \bmod 2 .
		\end{equation}
	We proceed by induction on~$\abs*{A_p(t)}$.
	If~$\abs*{A_p(t)} = 0$, then process~$p$
		is a local maximum among its active neighbors,
		and so in \Cref{alg:slow5} we have~$C^+ \gets \emptyset$,
		which implies~$a_p(t) = 0$.
	For the inductive step, suppose the result true for~$\abs*{A_p(t)} = k$,
		and suppose that~$\abs*{A_p(t)} = k+1$.
	Since process~$p$ is assumed to be non-minimal,
		it has one neighbor~$q$
		with~$\hat{X}_q(t) > \hat{X}_p(t)$
		and~$\abs{\hat{A}_q(t)} = k$,
		and we have~$a_p(t) = \min \Natural \setminus \Set{\hat{a}_q(t), \hat{b}_q(t)}$.

	If~$k$ is even,
		then by inductive assumption
		we have~$\hat{a}_q(t) = 0$,
		and so~$a_p(t) \neq 0$.
	Otherwise, $k$ is odd,
		and by inductive assumption
		we have~$\hat{a}_q(t) > 0$.
	In the code of \Cref{alg:slow5},
		we have~$C^+ \subseteq C$,
		and so for any process~$u \in [n]$
		and time~$\tau \geqslant 0$
		we have~$b_u(\tau) \geqslant a_u(\tau)$.
	Thus in particular
		we have~$\hat{b}_q(t) \geqslant \hat{a}_q(t) > 0$,
		and therefore~$a_p(t) = 0$.

	For the main claim,
		let~$\ell \coloneqq \abs*{A_p(t_1)}
		= \abs*{A_p(t_2)} = \abs*{A_p(t_3)} = \abs*{A_p(t_4)}$.
	If~$\ell$ is even,
		then~$a_p(t) = 0$ for all~$t \in \interval{t_1}{t_4}$.
	Reasoning as in \Cref{lemma:slow6:stubborn},
		if~$p$ still hasn't returned by time~$t_4$,
		then we have~$\abs*{A_p(t_3)} = \abs*{A_q(t_3)} - 1 =
		\abs*{A_{q'}(t_3)} + 1$
		without loss of generality.
	Then if neither~$q$ nor~$q'$ is activated
		in the interval~$\interval{t_3+1}{t_4}$,
		$p$ terminates by \Cref{lemma:slow6:obfree}.
	Otherwise, using again the fact
		that~$b_u(\tau) \geqslant a_u(\tau)$
		for any process~$u$ and time~$\tau$,
		we have~$\hat{a}_p(t_4) = 0 < \min \Set{\hat{a}_{q}(t_4),
		\hat{b}_{q}(t_4),\hat{a}_{q'}(t_4),\hat{b}_{q'}(t_4)}$,
		and so~$p$ returns in~$t_4$.

	If~$\ell$ is odd,
		we suppose without loss of generality
		that~$X_q > X_p > X_{q'}$.
	We have~$\hat{a}_p(\tau) > 0$
		for all~$\tau \in \interval{t_2}{t_4}$,
		and reasoning again as in \Cref{lemma:slow6:stubborn},
		by time~$t_4$ we have~$\hat{a}_{q'}(t_4) = 0$,
		and~$p$ terminates if it is still active.
\end{proof}

\begin{lemma}\label{lemma:slow5:complexity}
	Let $p\in[n]$ be a process
		that is \emph{not} a local minimum for the identifiers,
		and let~$\ell$ denote the monotone distance
		from~$p$ to the closest maximal process.
	Process~$p$ returns after at most~$3 \, \ell + 4$ activations.
\end{lemma}

\begin{proof}
	This is a direct consequence of the previous lemma:
		for~$p$ to keep working,
		its set~$A_p(t)$ must increase
		at least every~$4$ activations.
	The claim follows.
\end{proof}

\begin{proof}[Proof of \Cref{theo:alg:slow5}]
	Thanks to \Cref{lemma:slow5:complexity},
		processes that are \emph{not} local minima
		return after a number of steps
		that is at most~$\Floor{3n/2} + 4$.
	Local minima terminate at most  one step after their two neighbors have terminated,
		i.e., in at most~$3n + 8$ rounds.
	The proper coloring is an immediate consequence
	of \Cref{lemma:slow5:avoid}.
\end{proof}

\section{From Linear Time to Almost Constant Time}\label{section:fast5}

Here, we augment \Cref{alg:slow5}
	with a mechanism designed to reduce~$X_p$,
	initially set to the identifier of the process.
As the identifiers\footnote{%
	For simplicity, we continue to refer to~$X_p(t)$
	as process~$p$'s \enquote{identifier},
	even though it is now possible that~$X_q(t) = X_p(t)$
	for some other process~$q \nsim p$.}
	will now be evolving through time,
	we will say that a process~$p$, with neighbors~$q,q'$,
	is a local extremum at time \textit{$t \geqslant 1$}
	if~$\hat{X}_p(t) > \hat{X}_q(t),\hat{X}_{q'}(t)$.
The resulting algorithm, displayed as \Cref{alg:fast5},
	5-colors the cycle~$C_n$ in $\BigO{\log^* n}$ steps.

The intuition for \Cref{alg:fast5} is as follows.
Every process~$p$ essentially runs \Cref{alg:slow5} unchanged,
	and stops whenever this algorithm terminates.
However, in parallel, every process~$p$ updates its identifier~$X_p$,
	initially equal to the identifier of~$p$,
	\textit{à la} Cole and Vishkin
	using a reduction function~$f$ defined hereafter.
This helps to reduce long monotone chains of identifiers
	to a constant length,
	speeding up the convergence of \Cref{alg:slow5}.
This addition to the algorithm is blocking, as,
	to maintain a proper coloring of the identifiers~$X_p$
	(which is crucial for the wait-free coloration algorithm),
	every process~$p$ must wait for the approval
	of both its neighbors
	each time $p$ wants to update its identifier,
	through the use of a local counter~$r_p$
	which tracks the number of times process~$p$
	tried to pick a smaller identifier.
If all processes advance \enquote{almost synchronously},
	then they quickly (in $\BigO{\log^* n}$ steps)
	reach a stage where the remaining monotone chains of identifiers
	are all shorter than a constant~$L \leqslant 10$.
From then on, the algorithm behaves as \Cref{alg:slow5},
	and all processes terminate in~$\BigO{L}$ steps,
	that is, in constant time.
The crux of the proof is therefore to show that slow processes
	cannot delay the convergence of fast processes too much. 
Indeed, a slow process may delay other processes,
	but if it blocks them during too many iterations
	(with respect to the reduction of the identifiers~$X_p$),
	then the system starts behaving as \Cref{alg:slow5},
	and neighboring processes actually quickly terminate.
On the other hand,
	if a process is only \enquote{moderately slow},
	and allows its neighbors to make some progress on the reduction of their identifiers~$X_p$,
	then other processes use this property for breaking symmetry,
	and they stop waiting for the slow process.

\subsection{Reducing identifiers with deterministic coin-tossing}

The considerable speedup achieved in comparison to \Cref{alg:slow5}
	relies on an identifier-reduction function~%
	$f: \Natural \times \Natural \to \Natural$,
	adapted from Cole and Vishkin's algorithm~\cite{ColeV86},
	defined as follows.
For any natural number $Z$,
	we denote its binary decomposition by
	$Z = \sum_{k \in \Natural} Z_k 2^k$,
	whose length is
	$\abs{Z} \Def \Ceil*{\log_2(Z + 1)}$.
Given two natural numbers $X$ and $Y$,
	we then set
	\begin{equation}\label{eq:f-def}
		f(X,Y) = 2 \mkern1mu i + X_i \quad \text{where} \: i \Def
		\min \Set{\abs{X}, \abs{Y}} \cup \Set{ k \in \Natural \Where X_k \neq Y_k }
	\end{equation}

As~$f(x,y) \leqslant 2 \abs{x}+1 = \BigO{\log(x)}$,
	one reaches a constant fixed point
	after $\BigO{\log^* n}$ iterate calls to~$f$,
	which gives the following.
	Recall that, for $k \in \Natural$,  $k$-th iterate
		of a function $F : A \to A$
		is recursively defined
	as $F^{(0)}(x)=x$ and, for $k\geq 1$, $F^{(k)} = F \circ F^{(k-1)}$. 

\begin{lemma}\label{lemma:f-logstar}
	Let $F : \rinterval{1}{+\infty} \to \rinterval{1}{+\infty}$
		be the function~$x \mapsto F(x) = 2 \Ceil*{\log(x+1)} + 1$.
	There exists $\alpha>0$ such that,
	for every~$x \geqslant 1$,
	there exists $t\leq \alpha \log^* x$
	such that $F^{(t)}(x) < 10$.
\end{lemma}

\begin{lemma}\label{lemma:monotone-f}
	Let~$x, y \in \Natural$.
	If~$x > y \geqslant 10$, then $f(x,y) < y$.
\end{lemma}

\begin{proof}
	Let $\ell =\abs{y}$. By assumption, we have~$\ell\geqslant 4$.
	If~$\ell = 4$,
		then $f(x,y) \leqslant 2 \mkern1mu \ell + 1 = 9 < y$.
	If $\ell \geqslant 5$,
		then we have~$y \geqslant 2^{\ell-1}$, and so
		$y - f(x,y) \geqslant 2^{\ell - 1} - 2 \mkern1mu \ell - 1 > 0$, where 
		the last inequality is because~$2^z > 4 z + 2$
		whenever~$z \geqslant 5$.
\end{proof}

The proper coloring maintained by the function~$f$
	relies on the following Cole and Vishkin-like property.

\begin{lemma}\label{lemma:f-cole-vishkin}
	Let~$x,y,z \in \Natural$.
	If~$x > y > z$, then~$f(x,y) \neq f(y,z)$.
\end{lemma}

\begin{proof}
	Let~$f(x,y) = 2 i^* + x_{i^*}$.
	For all~$i < i^*$, $x_i = y_i$,
		and if~$i^* < \abs{y}$ then~$x_i \neq y_i$.
	Suppose that~$f(y,z) = f(x,y)$.
	Then~$y_{i^*} = x_{i^*}$,
		and by the above~$i^* \geqslant \abs{y} \geqslant \abs{z}$.
	In this case, $y_i = z_i$ for all~$i < \abs{y}$,
		and thus~$y = z$, contradicting our assumption~$y > z$.
\end{proof}

\subsection{5-coloring the cycle in near-constant time}

\begin{algorithm}
	\caption{Fast 5-coloring algorithm, code for process~$p$ with neighbors $q$ and $q'$}\label{alg:fast5}
		\Input{$X_p \in \Natural$}
\Initially{
		$a_p, b_p, r_p \gets 0 \in \Natural$\;
	}
	\Forever{
		\textbf{write}$(X_p, r_p, a_p, b_p)$ \text{and}
		\textbf{read}$((X_q, r_q, a_q, b_q),(X_{q'},r_{q'}, a_{q'}, b_{q'}))$\\ 
		\lIf{$a_p \notin \Set{a_q, b_q, a_{q'}, b_{q'}}$}{\Return{$a_p$}}
		\lElseIf{$b_p \notin \Set{a_q, b_q, a_{q'}, b_{q'}}$}{\Return{$b_p$}}
		\Else{
			$a_p \gets \min \Natural \setminus \Set{a_u, b_u \Where (u \sim p) \wedge (X_u > X_p)}$\;
			$b_p \gets \min \Natural \setminus \Set{a_q, b_q, a_{q'}, b_{q'}}$\;
			\If{$(r_p < \infty) \wedge (r_p \leqslant \min\Set{r_q, r_{q'}})$}{
				\uIf{$\min\Set{X_q,X_{q'}} < X_p < \max\Set{X_q,X_{q'}}$}{
					$r_p \gets r_p + 1$\;
					$Y \gets f(X_p, \min\Set{X_q,X_{q'}})$\Comment*{$f$ given
					in \Cref{eq:f-def}}
		
					\lIf{$Y < \min\Set{X_q,X_{q'}}$}{$X_p \gets Y$}
				}
				\Else{
					$r_p \gets \infty$\;
					\If{$X_p < \min \Set{X_q, X_{q'}}$}{
						$X_p \gets \min \Set*{X_p, \min (\Natural \setminus \Set{f(X_q,X_p), f(X_{q'},X_p)})}$
					}
				}
			}
		}
	}
\end{algorithm}

\begin{theorem}\label{theo:fast5}
	In any execution of \Cref{alg:fast5} over the cycle~$C_n$
		with a proper coloring
		provided by the values $(X_p)_{p\in [n]}$
		given to the processes as input:
		\begin{description}
			\item[\textsf{Termination:}]
				every process terminates
				after having been activated
				at most $\BigO{\log^* n}$ times;
			\item[\textsf{5-color palette:}]
				every process that terminates
				outputs a color in the set~$\Set{0, \ldots, 4}$;
			\item[\textsf{Correctness:}]
				the outputs properly color the graph
				induced by the terminating processes in~$C_n$.
		\end{description}
\end{theorem}

A crucial ingredient in the proof of correctness
	is to establish that the coloring provided
	by the evolving values of the local variables~$X_p$, $p\in [n]$,
	is always proper throughout any execution.

\begin{lemma}\label{lem:fast5:x-colored}
	Let~$p, q \in [n]$ be neighboring processes.
	For every~$t \in \Natural$, if
		$\hat{X}_p(t) \neq \bot$ then $\hat{X}_p(t) \neq \hat{X}_q(t)$.
\end{lemma}

\begin{proof}[Proof of~\Cref{lem:fast5:x-colored}]
	We show the following: for every~$t \in \Natural$,
		$X_p(t) \notin \Set{X_{q}(t), \hat{X}_q(t)}$,
		proceeding by induction.
	The case~$t = 0$ results from
		the initial proper coloring of the identifiers.

	For the induction,
		suppose the claim holds for~$t = 0, \ldots, T$.
	If~$p,q \notin \overline{\sigma}(T+1)$,
		then nothing changed, and
		the claim still holds for~$t=T+1$.

	Suppose~$p,q \in \overline{\sigma}(T+1)$.
	If~$r_q(T+1) = r_q(T)$, the claim immediately follows,
		as does it if~$X_q(T+1) = X_q(T)$.
	Otherwise, by assumption we either have~$X_q(T) > X_p(T)$, or the opposite.
	If the former, $X_q(T+1) = f(X_q(T), X_p(T)) < X_p(T)$,
		and by \Cref{lemma:f-cole-vishkin}
		we have~$X_q(T+1) \notin \Set{X_p(T), X_p(T+1)}$,
		and so~$X_p(T+1) \notin \Set{\hat{X}_q(T+1), X_q(T+1)}$.
	Otherwise, $X_q(T) < X_p(T)$;
		if~$q$ is a local minimum in~$T+1$,
		then~$X_q(T+1) \neq f(X_p(T), X_q(T))$,
		and the claim follows
		from~$X_p(T+1) \in \Set{X_p(T), f(X_p(T), X_q(T))}$.
	If~$q$ is \emph{not} a local minimum,
		then~$X_q(T+1) = f(X_q(T),Z) < Z$ for some~$Z < X_q(T)$;
		here again, the claim follows from \Cref{lemma:f-cole-vishkin}.
	
	Finally, suppose~$p \in \overline{\sigma}(T+1)$
		and $q \notin \overline{\sigma}(T+1)$.
	If~$X_p(T+1) = X_p(T)$,
		then the claim still holds.
	Otherwise, we have~$r_p(T) < r_p(T+1) \leqslant \infty$,
		and~$X_p(T+1) < X_p(T)$.
	Process~$p$ is then not a local maximum in~$T+1$,
		and the algorithm guarantees~$X_p(T+1) < \hat{X}_q(T+1)$.
	If~$X_q(T+1) = \hat{X}_q(T+1)$,
		and in particular if~$r_q(T+1) = \hat{r}_q(T+1)$,
		and the claim holds.

	Suppose then that~$r_q(T+1) < \hat{r}_q(T+1)$,
		and let~$t_0$ be the earliest time when~$r_q(t_0) = r_q(T)$,
		such that~$\hat{r}_q(t_0) = \hat{r}_q(T+1)$.
	Process~$q$ takes no steps in the interval~$\linterval{t_0}{T+1}$,
		and because~$r_q$ increases in~$t_0$,
		we have~$\hat{r}_q(t_0) \leqslant \hat{r}_p(t_0)$.
	Thus~$\hat{r}_q(T+1) \leqslant \hat{r}_p(t_0) \leqslant \hat{r}_p(T+1)$.
	Since~$r_p$ increases in~$T+1$, we have
		$\hat{r}_p(T+1) \leqslant \hat{r}_q(T+1)$, and thus
		\begin{equation*}
			\hat{r}_q(t_0) = \hat{r}_p(t_0) = \hat{r}_q(T+1) = \hat{r}_p(T+1),
		\end{equation*}
		i.e., $\hat{r}_p(t)$ is constant for~$t \in \interval{t_0}{T+1}$,
		and as a consequence, $\hat{X}_p(T+1) = \hat{X}_p(t_0)$.
	From here, we proceed as in the previous case:
		$X_q(T+1) = X_q(t_0)$ was computed
		with~$q$ seeing~$\hat{X}_p(t_0) = \hat{X}_p(T+1)$,
		and, since~$X_p(T+1)$ was computed
		with~$p$ seeing~$\hat{X}_q(T+1) = \hat{X}_q(t_0)$,
		we have indeed~$X_p(T+1) \notin \Set{X_q(T+1), \hat{X}_q(T+1)}$.
\end{proof}

When discussing executions of \Cref{alg:fast5},
	we say that a process~$p$ is ""blocked"" at time~$t$
	if~$r_p(t)< \infty$, $r_p(t) = \hat{r}_p(t)$,
	and~$p$ has not terminated at time~$t$.
Since the value of~$X_p$ changes only if~$r_p$ increases,
	we have~$X_p(t) = \hat{X}_p(t)$ whenever process~$p$ is blocked at time~$t$.
A process~$p$ that \emph{is not} blocked at time~$t$,
	will write a new value for~$\hat{r}_p(t)$ at its next activation.
Moreover, $p$~writes a new value for~$\hat{X}_p(t)$ as well,
	unless $p$ satisfies specific properties: $X_p$ is a local maximum,
	$X_p$ is a local minimum,
	or~$p$ has a neighbor~$q$ with~$\hat{X}_q < 10$.
Note that, before its first activation,
	every process~$p$ is unblocked,
	as~$r_p(0) = 0 \neq \hat{r}_p(0) = \bot$.

Every process that takes sufficiently many non-blocked steps,
	namely $\BigOmega{\log^* n}$ steps,
	quickly reduces its identifier~$X_p$
	until~$X_p$,
	or the identifier $X_q$ of one of its neighbors~$q$ becomes
	 smaller than~$10$.
At this stage of the execution,
	monotone chains of identifiers will cease to evolve
	after an additional constant number of steps.
Once the monotone chains of identifiers cease to evolve,
	the analysis developed in the previous section
	shows that processes terminate
	in a number of steps that is not larger
	than the length of monotone chains of identifiers,
	which is itself bounded by a constant~$L \leqslant 10$.
In other words,
	when all processes take~$\BigOmega{\log^* n}$ non-blocked steps,
	they terminate in an additional~$\BigO{1}$ steps.

In the following, we then focus on the case
	where the identifiers of the processes
	are still greater than~$10$,
	and we will show fast convergence is guaranteed
	even in the presence of blocked processes.
Indeed, the main difficulty in proving \Cref{theo:fast5}
	is to deal with blocked processes.
Mainly, we show that a process quickly terminates
	whenever it is not blocked at \emph{too many} steps.
	
\begin{lemma}\label{lemma:max-forever}
		Let $p\in[n]$ be a process.
		For all~$t \geqslant 1$,
			if~$p \in \overline{\sigma}(t)$
			and~$\hat{X}_p(t)$ is a local maximum in some time~$t$,
			then~$\hat{X}_p(t')$ is a local maximum for all~$t' \geqslant t$.
\end{lemma}

\begin{proof}
	Local maxima never update their identifiers;
		other processes might, but only to decrease them.
	The claim follows.
\end{proof}

\begin{lemma}\label{lemma:blocked-max}
	Let~$p\in[n]$ be a process, and let $q,q'$ be its two neighboring processes in the cycle.
	Let us assume that processes~$p$ and~$q$ are blocked at some time~$t_0$,
		with~$\hat{r}_q(t_0) < \hat{r}_p(t_0)$, $\hat{X}_q(t_0) > \hat{X}_p(t_0) > \hat{X}_{q'}(t_0)$,
		and~$\hat{X}_p(t_0) \geqslant 10$.
	Additionally, let us assume that process~$q$
		remains blocked throughout the whole time interval~$\rinterval{t_0}{t_1}$ for some $t_1>t_0$,
		but becomes activated and unblocked at time~$t_1$.
Then, one of the following claims holds: 
\begin{itemize}
	\item $X_q$ is a local maximum at time~$t_1$.
	\item If process $q$ is activated again at some time~$t_2 > t_1$,
		then~$X_p$ will become a local maximum
		as soon as $p$ is activated at a time~$t \geqslant t_2$.
\end{itemize}
\end{lemma}

\begin{proof}[Proof of~\Cref{lemma:blocked-max}]
	Since~$p \sim q$, and since~$p$ is blocked at time~$t_0$,
		we have~$\hat{r}_p(t_0) = \hat{r}_q(t_0)+1$.
	Moreover, as~$\hat{X}_p(t_0) \geqslant 10$,
		the fact that~$\hat{X}_p(t_0)$ is not a local minimum
		means that~$\hat{r}_{q'}(t_0) \geqslant \hat{r}_p(t_0)$. Otherwise, by \Cref{lemma:monotone-f}, $X_p$ would be smaller than $X_{q'}$.
	In particular, process~$p$ remains blocked
		as long as process~$q$ is itself blocked.

	Now, suppose that~$r_q(t_1) \neq \hat{r}_q(t_1)$.
	As processes~$p,q$ are blocked until~$t_1$,
		we have~$\hat{X}_q(t_1) = \hat{X}_q(t_0)$ and~$\hat{X}_p(t_1) = \hat{X}_p(t_0)$,
		so~$\hat{X}_q(t_1) > \hat{X}_p(t_1)$,
		and~$q$ is not a local minimum in~$t_1$.
	The case~$r_q(t_1) = \infty$
		thus corresponds to~$\hat{X}_p(t_0)$ being a local maximum at time~$t_1$.
	If $r_q(t_1) < \infty$, then $r_q(t_1) = \hat{r}_q(t_0)+1$,
		and since~$\hat{X}_p(t_1) = \hat{X}_p(t_0) \geqslant 10$, we get $X_q(t_1) = f(\hat{X}_q(t_0),\hat{X}_p(t_0)) < \hat{X}_p(t_0)$ by \Cref{lemma:monotone-f}.

	Finally, suppose then that~$q$ is next activated at time~$t_2$,
		and that~$p$ is activated at time~$t \geqslant t_2$.Note that as long as $q$ does not get activated again, $p$ remains blocked, as it did not see the update on $r_q$. Moreover, as $X_q(t_1)$ is not a local maximum, $X_q(t_2)<\hat{X}_p(t_1)=\hat{X}_p(t_0)$.
	Then process~$p$ sees~$\hat{X}_p(t_0)$ to be a local maximum,
		and since it is no longer blocked by~$q$
		we have~$r_p(t) = \infty$.
\end{proof}

\begin{lemma}\label{lemma:blocked-linear}
	Let $k\geq 1$, and let $q_0 \sim q_1 \sim \cdots \sim q_k$
		be a sequence of $k+1$ distinct processes in the cycle.
	Let $t_0 \in \Natural$,
		and let $t_1 \in \Natural \cup \Set{\infty}$ with $t_1>t_0$.
	Let us assume that
		(1)~for every~$t \in \rinterval{t_0}{t_1}$,
		$q_0 \notin \overline{\sigma}(t)$,
		(2)~processes $q_1, \ldots, q_k$ are "blocked" at time~$t_0$,
		with~$\hat{r}_{q_0}(t_0) < \hat{r}_{q_1}(t_0) < \cdots < \hat{r}_{q_k}(t_0) < \infty$,
		and (3)~$\hat{X}_{q_k}(t_0) \geqslant 10$.
	Then, for every $i \in \Set{1,\dots,k}$,
		process~$q_i$ terminates
		after having been activated at most~$3i+1$ times
		in the time interval~$\rinterval{t_0}{t_1}$.
\end{lemma}

\begin{proof}[\Cref{lemma:blocked-linear}]
	Under the hypotheses of the lemma,
		processes~$q_1, \ldots, q_k$ remain blocked
		throughout the time interval~$\interval{t_0}{t_1-1}$,
		and we have~$\hat{X}_{q_0}(t) > \hat{X}_{q_1}(t) > \cdots > \hat{X}_{q_k}(t)$
		whenever~$t_0 \leqslant t < t_1$.
	By the same arguments as the ones used in the previous section
		for establishing \Cref{lemma:slow6:parity},
		the sign of~$a_{q_i}(t)$
		is determined by the sign of the variables~$a_{q_0}, \ldots, a_{q_{i-1}}$
		throughout the time interval~$\interval{t_0}{t}$.
	In particular, the sign of~$\hat{a}_{q_i}(t)$ stabilizes
		after $q_i$ has been activated at most~$3\,i$ times,
		and thus process~$i$ itself terminates after having been activated
		at most~$(3i+1)$ times.
\end{proof}

\begin{lemma}\label{lemma:near-max}
	Let $p, q \in [n]$ be two neighboring processes.
	If~$X_q$ is a local maximum at time~$t_0 \in \Natural$,
		and if~$\hat{r}_q(t_0) = \infty$,
		then~$p$ terminates
		after having been activated $\BigO{\log^* n}$ times
		during the time interval~$\rinterval{t_0}{\infty}$.
\end{lemma}

\begin{proof}[Proof of~\Cref{lemma:near-max}]
	Let~$t_1, t_2, \ldots$ be the consecutive steps
		taken by process~$p$ in the interval~$\rinterval{t_0}{\infty}$,
		that is, for every~$k \geqslant 1$,
		$p$ is inactive during the whole time~$\interval[open]{t_k}{t_{k+1}}$.
	Note that, since~$\hat{r}_q(t_0) = \infty$,
		we have $\hat{a}_q(t) = 0$ for all~$t \geqslant t_0$,
		and so $\hat{a}_p(t) > 0$ for all~$t \geqslant t_2$,
		as process~$q$ will remain a local maximum forever.

	Pick~$k \geqslant 2$.
	If~$p$ is \emph{not} a local minimum
		at any point in~$\interval{t_k}{t_{k+3}}$,
		then by \Cref{lemma:slow5:Ap} it terminates in~$t_{k+3}$ at the latest.
	Conversely, if~$p$ \emph{is} a local minimum
		throughout the same interval,
		then we have repeatedly~$\hat{a}_p(t_i) \in \Set{\hat{a}_{q'}(t_i),
		\hat{b}_{q'}(t_i)}$,
		$i = k, \ldots, k+3$.
	This implies that~$\hat{a}_{q'}$
		is positive during that interval,
		otherwise $p$ and~$q'$ would eventually stop having conflicts.
	By the same argument, process~$q'$
		terminates in a constant number of activations,
		and so do process~$p$.
	Therefore, for process~$p$ \emph{not} to terminate
		the relative order of~$\hat{X}_p$ and~$\hat{X}_{q'}$
		must switch every~$\BigO{1}$ steps.
	Thus, every time~$p$ takes~$\BigO{1}$ steps and fails to return,
		it must be the case that~$\hat{X}_p$ has decreased.
	As argued before, this can happen at most~$\BigO{\log^* n}$ times
		before either~$X_p \leqslant 10$ or~$X_{q'} \leqslant 10$,
		at which point convergence happens in a bounded number of steps.
\end{proof}

\begin{lemma}\label{lemma:blocked-logstar}
	Let $p \in [n]$ be a process.
	If $p$ is "blocked" at every time~$t\in\rinterval{t_0}{t_1}$,
		and if~$p$ takes~$4$ steps during that interval,
		then~$p$ takes up to~$\BigO{\log^* n}$ steps in~$\rinterval{t_0}{\infty}$
		before terminating.
\end{lemma}

\begin{proof}[Proof of~\Cref{lemma:blocked-logstar}]
	Since process~$p$ is "blocked",
		a direct induction shows that~$p$
		lies somewhere within a monotone chain of identifiers,
		as described in \Cref{lemma:blocked-linear}.
	That is, there is a chain of distinct adjacent processes
		\begin{equation*}
		q_{-k-1} \sim q_{-k} \sim \cdots \sim q_{-1}
		\sim q_0 \sim q_{1} \sim \cdots \sim q_{\ell},
		\end{equation*}
		with $q_0=p$, and $k, \ell \geqslant 0$, where,
		for every $i \in \interval{-k}{\ell}$,
		$\hat{r}_{q_i}(t_0) = \hat{r}_p(t_0) + i$,
		and either~$\hat{r}_{q_{-k-1}}(t_0) = \bot$
		(in which case $k = \hat{r}_p(t_0)$)
		or~$\hat{r}_{q_{-k-1}}(t_0) = R-k-1$
		(in which case $k < \hat{r}_p(t_0)$.
	Moreover, all processes $q_{-k}, \ldots, q_{\ell}$
		are "blocked" at time~$t_0$,
		and process~$q_{-k-1}$ is not "blocked" at time~$t_0$.

	We now distinguish two cases, depending on whether~$k = 0$ or not.
	If~$k = 0$, then, by \Cref{lemma:blocked-linear},
		process~$p$ terminates after taking~$4$ steps
		within the time interval~$\rinterval{t_0}{t_1}$.
	If $k > 0$, then process~$q_{-1}$ is "blocked";
		if process~$q_{-1}$ remains "blocked" while process~$p$ takes~$3k+1$ steps,
		then, by \Cref{lemma:blocked-linear}, $p$ terminates.
	The same holds if~$q_{-1}$ takes a single non-blocked step.
	If process~$q_{-1}$ ever becomes unblocked, and takes another step,
		then we meet the assumptions of \Cref{lemma:blocked-max},
		and either of~$X_p$ or~$X_{q_{-1}}$ become a local maximum.
	If~$X_p$ becomes a local maximum, then it terminates in~$\BigO{1}$ steps.
	If~$X_{q_{-1}}$ becomes a local maximum,
		then, by \Cref{lemma:near-max},
		process~$p$ terminates in~$\BigO{\log^* n}$ steps.
\end{proof}

We are now ready to show \Cref{theo:fast5}.

\begin{proof}[Proof of \Cref{theo:fast5}]
	For a process $p$, there are two possible paths,
		both leading to $p$ returning:
	\begin{enumerate}
		\item Process $p$ never gets blocked.
			By \Cref{lemma:f-logstar},
				if a process updates its identifier
				up to~$\BigO{\log^* n}$ times,
				its identifier ends up in the interval~$\interval{0}{10}$.
			Therefore, after $\BigO{\log^*n}$ rounds,
				either $X_p$ becomes a local maximum,
				or $X_p \leqslant 10$.
			In the first case,
				it stays a maximum by \Cref{lemma:max-forever},
				its $a_p(t)$ remains constant,
				and $p$ terminates after 4 rounds,
				thanks to \Cref{lemma:slow5:Ap}.
			In the second case, it will stay at distance
				at most~$10$ from a local minimum.
			As the processes of this path will no longer change their $X_{{-}}$,
			\Cref{lemma:slow5:complexity} allows us to conclude.
		\item Process $p$ becomes blocked.
			This can happen after at most $\BigO{\log^*n}$ rounds
				(otherwise we would end up in the previous case).
			\Cref{lemma:blocked-logstar} ensures that
				at most $\BigO{\log^*n}$ rounds will happen
				before $p$ returns.
	\end{enumerate}
This complete the proof that 5-coloring the asynchronous cycles $C_n$, $n\geq 3$,
	can be achieved in $O(\log^*n)$ rounds. 
\end{proof}

\section{Conclusion and future works}

We have presented a wait-free distributed algorithm
for proper coloring the $n$ nodes of the asynchronous cycle $C_n$, for every $n \geq 3$.
This algorithm performs in $\BigO{\log^*n}$ rounds, which is optimal, thanks to Linial's lower bound~\cite{Linial92} that applies even to the synchronous execution.
The algorithm uses 5~colors to proper color any cycle~$C_n$, $n\geq 3$, which is equal to the minimum number~5 of colors required to proper color the asynchronous cycle~$C_3$~\cite{AttiyaP16,CastanedaR10,HerlihyS99}.
Even if, for $n>3$, the existence of a 3-coloring algorithm is not directly ruled out by~\cite{AttiyaP16,CastanedaR10,HerlihyS99}, we conjecture that $k$-coloring the $n$-node cycle~$C_n$ requires $k\geq 5$ for every $n\geq 3$.

A natural extension of this work is to consider wait-free coloring arbitrary graphs.
Note that, by the renaming lower bound,
	coloring graphs with maximum degree~$\Delta$
	requires a palette of at least $2\Delta+1$ colors
	whenever $\Delta+1$ is a power of a prime.
This is because the shared memory model
	and the model in this paper coincide
	in the case of coloring the clique of $n=\Delta+1$ nodes.
We do not know if $2\Delta+1$ colors
	suffice for properly coloring all graphs
	of maximum degree~$\Delta$ in a wait-free manner.
It is however easy to extend \Cref{alg:slow6} to graphs with maximum degree~$\Delta$, for every $\Delta\geq 2$, using a range of colors of size $O(\Delta^2)$ (see Appendix~\ref{app:coloring-general-graphs}).
The running time of this algorithm may however be as large as the one of \Cref{alg:slow6}, i.e., $O(n)$ rounds.
In the synchronous setting, there is an algorithm for $O(\Delta^2)$-coloring performing in $O(\log^*n)$
rounds~\cite{Linial92} in any graph.
However, the techniques used in the synchronous setting for reducing the number of colors from $O(\Delta^2)$ to $\Delta+1$  (see \cite{Maus21}) seems hard to transfer to the asynchronous setting.

More generally, it would be interesting to characterize which of classical graph problems studied in synchronous failure-free networks can be solved wait-free in asynchronous networks, and which cannot.
And, for those solvable wait-free,
	what are their round-complexities?
For instance, 5-coloring can be solved wait-free
	in the asynchronous cycle, in $O(\log^*n)$ rounds,
	while maximal independent set (\MIS{})
	cannot be solved at all in asynchronous cycles.

\bibliography{references}

\appendix

\section{Coloring General Graphs}
\label{app:coloring-general-graphs}

It is possible to extend \Cref{alg:slow6}
	to connected graphs with maximum degree~$\Delta$,
	for every $\Delta\geq 2$ (see \Cref{alg:deltasquare}).
By construction, every process running \Cref{alg:deltasquare}
	returns a colors taken in the set
	\begin{equation*}
		\Set{(a,b) \Where a + b \leqslant \Delta},
	\end{equation*}
	of cardinality~$\frac{(\Delta+1)(\Delta+2)}{2}=\BigO{\Delta^2}$.
The proof of correctness for \Cref{alg:deltasquare}
	uses the same arguments
	as for establishing the correctness of \Cref{alg:slow6}.
In particular, a process cannot run forever
	whenever its identifier becomes a local extremum
	among the identifiers of its active neighbors,
	which guarantee that every process eventually terminates.

\begin{algorithm}
	\caption{$\BigO{\Delta^2}$-coloring algorithm for general graphs, code for process~$p$ with neighbors $q_1,\dots,q_k$, $k\leq \Delta$.}\label{alg:deltasquare}
	\Input{$X_p \in \Natural$}
	\Initially{
		$c_p = (a_p, b_p) \gets (0,0) \in \Natural \times \Natural$
	}
	\Forever{
		\textbf{write}$(X_p, c_p)$ \text{and} \textbf{read}$((X_{q_1}, c_{q_1}),\dots,(X_{q_k}, c_{q_k}))$\Comment{immediate snapshot}
		\lIf{$c_p \notin \Set{c_{q_1}, \ldots, c_{q_k}}$}{\Return{$c_p$}}%
		\label{alg:deltasquare:return}
		\Else{
			$a_p \gets \min \Natural
			\setminus \Set{a_u \Where u \sim p, X_u > X_p}$
			\label{alg:deltasquare:alpha}\;
			$b_p \gets \min \Natural
			\setminus \Set{b_u \Where u \sim p, X_u < X_p}$
			\label{alg:deltasquare:beta}\;
		}
	}
\end{algorithm}

\end{document}